\documentclass[12pt]{article}
\usepackage{epsfig}

\parskip=\medskipamount 
plus.3em minus.5em \abovedisplayshortskip=.5em plus.2em minus.4em
\renewcommand{\thesection}{\arabic{section}}

\catcode`@=11

\@addtoreset{equation}{section} \@addtoreset{equation}{subsection}
\def\theequation{\ifnum\value{section}=0 \arabic{equation}\ignorespaces
\else \ifnum\value{section}=-1 A.\arabic{equation}\ignorespaces
\else \ifnum\value{subsection}=0
\thesection.\arabic{equation}\ignorespaces \else
\thesection.\arabic{subsection}.\arabic{equation}\ignorespaces
                             \fi
                        \fi
                   \fi}

{\catcode`\'=\active \def'{{}^\bgroup\prim@s}}

\catcode`@=12

\newcommand{\bq}{\begin{equation}}
\newcommand{\be}{\begin{equation}}
\newcommand{\fq}{\end{equation}}
\newcommand{\ee}{\end{equation}}
\newcommand{\bqr}{\begin{eqnarray}}
\newcommand{\beqs}{\begin{eqnarray}}
\newcommand{\fqr}{\end{eqnarray}}
\newcommand{\eeqs}{\end{eqnarray}}

\newcommand{\rf}[1]{(\ref{#1})}

\def\bop#1{\setbox0=\hbox{$#1M$}\mkern1.5mu
    \vbox{\hrule height0pt depth.04\ht0
    \hbox{\vrule width.04\ht0 height.9\ht0 \kern.9\ht0
    \vrule width.04\ht0}\hrule height.04\ht0}\mkern1.5mu}

\begin{document}
\thispagestyle{empty}

\begin{flushright}
\begin{tabular}{l}
hep-th/0505077 \\
\end{tabular}
\end{flushright}

\vskip .6in
\begin{center}

{\bf Quantum Gauge Theory Amplitude Solutions}

\vskip .6in

{\bf Gordon Chalmers}
\\[5mm]

{e-mail: gordon@quartz.shango.com}

\vskip .5in minus .2in

{\bf Abstract}

\end{center}

The $n$-point amplitudes of gauge and gravity theory are given as a 
series in the coupling.  The recursive derivative expansion is used 
to find all of the coupling coefficients.  Initial conditions 
to any bare Lagrangian, or of an improved action, are used to compute 
quantum amplitudes.  

\vfill\break

\noindent {\it Introduction}

Gauge theory amplitudes and correlators have been well studied for 
many years.  Many techniques have been developed so as to compute 
tree and one-loop amplitudes, and holographic duals to supersymmetric 
gauge theories also gave computational tools.  The amplitude and 
correlator calculations in these theories, either in weak coupling 
or strong coupling, are typically given to low orders.  

There are a variety of methods to extend the scope of these calculations, 
and to find complete expressions for amplitudes and correlators.  This 
is a means in quantum field theory to find order unity, or large coupling 
results. 

The amplitudes of pure gauge theories are given here as a series in the  
coupling, $g$ or $G_N$.  The derivative expansion has been formulated and 
applied to many theories \cite{Chalmers4}-\cite{Chalmers13}.  The solution 
to this iteration is presented for gauge theory models; the scalar field 
theory coupling expansion is given in \cite{Chalmers3}.  There is a further 
mathematical simplification possible within the tensor formulation in the 
contributions.  

The Lagrangians considered are, 

\bqr 
{\cal L}_{YM} = \int d^dx ~{1\over g^2} {\rm Tr} F^2 + 
\sum {\cal O(F,\nabla)}_i  
\fqr 
with the summation on all the possible higher dimension operators (e.g. 
so-called irrelevant due to their perturbative scaling).  The gravitational 
theories are, 

\bqr 
{\cal L}_{G} = \int d^dx\sqrt{g} ~{1\over G_N} R + 
\sum {\cal O(R,\nabla)}_i \ , 
\fqr 
with a possibly infinite tower of higher dimension operators.   The 
classical scattering of the $F^2$ and $R$ actions are given in 
\cite{Chalmers1} (those in scalar theory in \cite{Chalmers2}; 
the number basis representation is very useful for labeling all of the 
contributions, including tree graphs containing ghosts.  These tree 
graphs and the numbers parameterizing them enter into the quantum 
solution through the initial conditions.  

The derivative expansion is used to find the perturbative quantum 
amplitudes.  Instantons are not included; their momentum structure 
is expected to have a number basis as do the classical scattering.

\vskip .2in 
\noindent {\it Amplitudes}

The coupling expansions of scalar field theory amplitudes have appeared 
in \cite{Chalmers1} and contain the required integral and tensor 
calculations.  The primary complications in the gauge theory calculation 
is due to the masslessness and tensor algebra.  

The recursive approach to the amplitude calculations has a solution 
presented in \cite{Chalmers3}.  The 
latter is represented in Figure 1 with the rainbow graphs; a sum of 
these graphs with any number of nodes is neccesary, and with internal 
ghost lines in covariant gauges, subject to the 
coupling addition that $q=\sum_{b=1}^{b_{\rm nodes}} q_b$ ($q_b$ is 
the coupling power of classical scattering at node $b$ which is 
$n-2$ for Yang-Mills theory).

The classical scattering is required to specify the tensors at the nodes.  
The ${\rm Tr} F^2$ Yang-Mills scattering can be obtained in a well ordered 
manner with the use of string theory.

The $\kappa(a;1)$ and $\kappa(b;2)$ set of primary numbers used on 
the string inspired set of Greens functions numbers produces the 
contributions, 

\bqr 
(-{1\over 2})^{a_1} ({1\over 2})^{n-a_2} 
\prod_{i=1}^{a_1} \varepsilon(\kappa(i;1))\cdot \varepsilon(\kappa(i;1))  
\times \prod_{j=a_1+1}^{a_2} \varepsilon(\kappa(j;1)) \cdot k_{\kappa(j;2)} 
\fqr 
\bqr 
\times \prod_{p=a_2+1}^n k_{\kappa(p;1)} \cdot k_{\kappa(p;2)}  \ ,
\label{kappaterms}
\fqr 
together with the permutations of $1,\ldots,n$.  The permutations extract 
all possible combinations from the set of terms in the labeled $\phi^3$ 
diagram, after distributing the numbers into the three categories.   

The form of the amplitudes are expressed as, 

\bqr
{\cal A}^n_\sigma = \sum_\sigma 
 C_\sigma g^{n-2} T_\sigma \prod t_{\sigma(i,p)}^{-1} \ , 
\label{gaugeamps}
\fqr 
with $T_\sigma$ in \rf{kappaterms} derived from the tensor set of 
$\kappa$, e.g. found from $\phi_n$ or the momentum routing of the propagators 
with $\sigma(i,p)$.   The normalization is $i(-1)^n$.  The numbers 
$a_1$ and $a_2$ are summed so 
that $a_1$ ranges from $1$ to $n/2$, with the boundary condition 
$a_2\geq a_1+1$.  Tree amplitudes in gauge theory must possess at least 
one $\varepsilon_i\cdot\varepsilon_j$.

All $\phi^3$ diagrams are summed at $n$-point, which is represented by the 
sum in $\sigma$ in \rf{gaugeamps}.  The color structure is 
${\rm Tr} \left(T_{a_1}\dots T_{a_n}\right)$,  and the complete amplitude 
involves summing the permutations of $1,\ldots, n$.  

The first $n-2$ numbers in $\kappa_2$ are summed beyond those of the primary 
numbers in accord with the set $i$ to $i+p-1$ for a given vertex label 
$i+p-1$, which labels the vertex in $\phi_n$.

The propagators are in correspondence with $\phi^3$ diagrams, 

\bqr 
D_\sigma = g^{n-2} \prod {1\over t_{\sigma(i,p)} - m^2} \ . 
\label{phi3diagrams} 
\fqr 
The Lorentz invariants $t_{\sigma(i,p)}$ are defined by $t_i^{[p]}$, 
 
\bqr  
t_i^{[p]} = (k_i+\ldots + k_{i+p-1})^2 \ .  
\label{invariants}
\fqr 
Factors of $i$ in the propagator and vertices are placed into the prefactor of 
the amplitude.
The sets of permutations $\sigma$ are what are required in order to specify 
the individual diagrams.  The full sets of $\sigma(i,p)$ form all of the 
diagrams, at any $n$-point order.  

The propagators are in correspondence with $\phi^3$ diagrams, 

\bqr 
D_\sigma = g^{n-2} \prod {1\over t_{\sigma(i,p)} - m^2} \ . 
\label{scalardiagrams} 
\fqr 
The Lorentz invariants $t_{\sigma(i,p)}$ are defined by $t_i^{[p]}$, 
 
\bqr  
t_i^{[p]} = (k_i+\ldots + k_{i+p-1})^2 \ .  
\label{momentainv}
\fqr 
Factors of $i$ in the propagator and vertices are placed into the prefactor of 
the amplitude.
The sets of permutations $\sigma$ are what are required in order to specify 
the individual diagrams.  The full sets of $\sigma(i,p)$ form all of the 
diagrams, at any $n$-point order.

The numbers $\phi_n(i)$ can be arranged into the numbers $(p_i,[p_i])$, in which 
$p_i$ is the repetition of the value of $[p_i]$.  Also, if the number $p_i$ equals 
zero, then $[p_i]$ is not present in $\phi_n$.  These numbers can be used to 
obtain the $t_i^{[q]}$ invariants without intermediate steps with the momenta.  
The branch rules are recognizable as, for a single $t_i^{[q]}$,  

\vskip .2in
\hskip .3in 0) $l_{\rm initial}=[p_m]-1$

\hskip .3in 1) 

\hskip .5in $r=1$ to $r=p_m$  

\hskip .5in ${\rm if~} r + \sum_{j=l}^{m-1} p_j = [p_m]-l_{\rm initial}   \quad {\rm then}~i= [p_m] 
  \quad q= [p_m] - l_{\rm initial}+1$ 

\hskip .5in beginning conditions has no sum in $p_j$

\hskip .3in 2) 

\hskip .5in ${\rm else~} \quad l_{\rm initial}\rightarrow l_{\rm initial}-1$ : 
decrement the line number

\hskip .5in $l_{\rm initial}>[p_{l}]$ else $l\rightarrow l-1$ : decrement the $p$ sum 

\hskip .3in 3) ${\rm goto}~ 1)$ 
\bqr  
\label{branchrules}
\fqr  
The branch rule has to be iterated to obtain all of the poles. 
This rule checks the number of vertices and matches to compare if 
there is a tree on it in a clockwise manner.  If not, then the external 
line number $l_{initial}$ is changed to $l_{initial}$ and the tree is 
checked again.  The $i$ and $q$ are labels to $t_i^{[q]}$.

\begin{figure}
\begin{center}
\epsfxsize=12cm
\epsfysize=12cm
\epsfbox{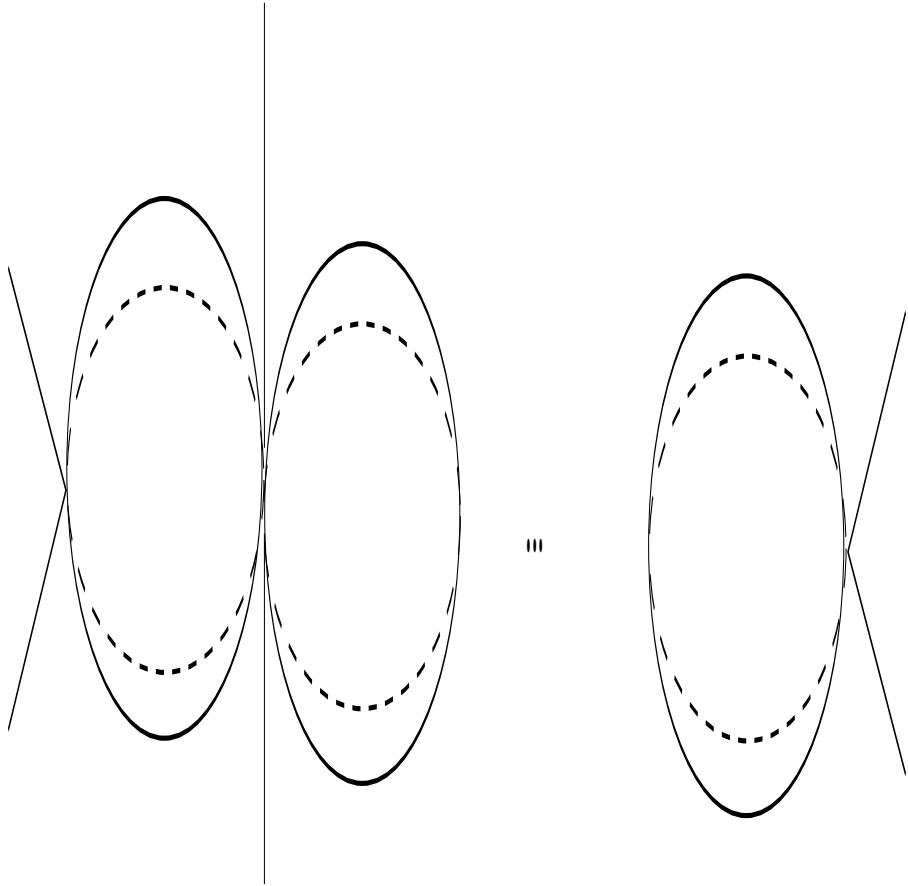}
\end{center}
\caption{ The product form solution to the recursive formulae defining 
the loop expansion.  The nodes are classical scattering vertices.}  
\end{figure}

\vskip .2in 
\noindent {\it Coupling Coefficients}

The recursive solution to the coupling coefficients are calculated in 
this section.  The solution to the recursive formulae is represented 
in Figure 1, and the graphs do not show the permutations on the external 
lines.  These diagrams were evaluated in scalar theory in \cite{Chalmers3}.

The tensor integrals between two adjacent nodes are, 

\bqr 
\int d^dx e^{ik\cdot x} \prod^n \partial_{\mu_j} 
 \Delta(m,x)^N = T_{\mu_j}^n \sum_{a=1}^\infty \delta(N,a) 
  (m^2)^{a-N\beta_2/2} (k^2)^{-d/2-N(\beta_1-\beta_2/2)-a-n}  
\fqr 
 
\bqr 
= \gamma i^n \sum_{p=0}^\infty   
   {1\over p!} \sum_{p=1}^m \sum_{p_a} {}_{\vert_{\sum a p_a = p}}    
   {p!\over \prod_{p=1}^m (a p_a)!}  
 {\Gamma(N+1)\over\Gamma(N-m)} ~c_{p_a}
\fqr 
\bqr 
\times ~\rho(\beta_1,\beta_2) 
{\Gamma\bigl(d/2+N(\beta_1-\beta_2/2)+1\bigr)\over 
                      \Gamma\bigl(d/2+N(\beta_1-\beta_2/2)+1-a\bigr)}  
\fqr 
\bqr
\rho \sum_{l=0}^n {n!\over l!(n-l)!} (2-d)^{n-l} 
 {\Gamma(a-N\beta_2/2+1)\over\Gamma(a-N\beta_2/2+1-l)} (m^2)^{a-N\beta_2/2} 
 (k^2)^{-d/2-N(\beta_1-\beta_2/2)-a}  
\fqr 
\bqr 
\rho(\beta_1,\beta_2) 
{\Gamma\bigl(-N(\beta_1-\beta_2/2)+1-a\bigr) \over 
                      \Gamma\bigl(-N(\beta_1-\beta_2/2)+1\bigr)}
{\Gamma\bigl(-d/2-N(\beta_1-\beta_2/2)+1\bigr)\over 
                      \Gamma\bigl(-d/2-N(\beta_1-\beta_2/2)+1-a\bigr)}  
\fqr 
\bqr 
\times ~ {\Gamma\bigl(d/2+N(\beta_1-\beta_2/2)+a+1-n\bigr)\over
 \Gamma\bigl(d/2+N(\beta_1-\beta_2/2)+a+1\bigr)}
\fqr 
\bqr 
\sum_{\sigma_w,\tilde\sigma_w} \prod_{i=1}^w 
  \eta_{\mu_{\sigma(i)}\mu_{\tilde\sigma(i)}} 
    \prod_{i=1}^{n-w} k_{\mu_{\rho(i)}} ~  
  2^{-w} {\Gamma(-d/2-N(\beta_1-\beta_2/2)-a-n+1)\over 
       \Gamma(-d/2-N(\beta_1-\beta_2/2)-a-2n+1+w/2)}
\label{winintegrals}
\fqr 
The sums on $l$ and $a$ should be performed, in order to have a simplified 
expression at fixed tensor structure.  The momentum $k$ is the sum of the 
momenta on the exterior of the integral, i.e. $k=\sum_{j=1}^q k_j$.

General massless integrals in the gauge theory are more complicated than 
those in \rf{winintegrals} because the tree amplitudes have massless poles.  
The integral contains the two node factors, 

\bqr
\prod_{a=b,b+1} \prod_{\sigma_a(i,p)} t_{\sigma_a(i,p)}^{-1} \ , 
\label{vertexfactors}
\fqr 
with a series of propagators and derivatives, 

\bqr
\bigl(\prod_{j=1}^a \partial_{\mu_j} \bigr) \Delta^N  \ .
\label{tensorprops}
\fqr
The integrals include the singular terms in \rf{vertexfactors}.  The 
node momentum is defined by $P_b=\sum_i k_i$, with $i=\sigma(m_i-m_b)$ 
to $i=\sigma(m_f+{\tilde m}_b)$.  The nodal momentum spans the lines 
that are integrated; the indices also specify which invariants in 
\rf{vertexfactors} are internal-external and external-external to the 
loop. 

The vertex tensors are, 
\bqr  
(-{1\over 2})^{a_1} ({1\over 2})^{n-a_2} 
\prod_{i=1}^{a_1} \varepsilon(\kappa_a(i;1))\cdot \varepsilon(\kappa_a(i;1))  
\times \prod_{j=a_1+1}^{a_2} \varepsilon(\kappa_a(j;1)) \cdot k_{\kappa_a(j;2)}
\fqr 
\bqr 
\times \prod_{p=a_2+1}^n k_{\kappa_a(p;1)} \cdot k_{\kappa_a(p;2)} \ . 
\label{vertextensor}
\fqr 
The polarizations satisfy the on-shell identity 

\bqr 
\sum_{\lambda=\pm} \varepsilon_{\lambda,\mu} \varepsilon_{-\lambda,\nu} 
 = -\eta_{\mu\nu} + {k_\mu p_\nu + p_\nu k_\mu \over p\cdot k} \ . 
\fqr  

The propagators can be used in conjunction with a massless form of the 
integral in \rf{winintegrals} with the idenity, 

\bqr 
t_i^{[p]} = (k_i + \ldots + k_{i+p-1})^2 
\fqr 
\bqr 
= (k_i+\ldots+k_j + 
 {1\over N_1}(i\partial_k)^{-1})^2  
\fqr 
with the $N_1$ due to the fact that the derivative is taken on the 
$N_1$ propagators $\Delta^{N_1}$; these propagators are with the 
$N_1$ momenta in the invariant.  There is a $(2-d)$ scaled into the 
number $N_1$ so that the result $(2-d)$ times an integer.  The propagator 
satisfies the identity 

\bqr  
\partial_\mu \Delta = (2-d) {x_\mu\over x^2} \Delta
= (2-d) ({i\over\partial^k_\mu}) \Delta \ .  
\label{propderivative}
\fqr 
The derivatives in the set of propagators in \rf{tensorprops} 
compound the complexity.  The $i$ and $p$ are used in conjunction 
with the set of node momenta $k_a$ for $a=\sigma(m_i)$ to 
$a=\sigma(m_f)$, to determine the derivatives in 
\rf{propderivatives}.    

The massless integrals are 

\bqr 
\int d^dx e^{ik\cdot x} \Delta^N = 
   \int d^dx e^{ik\cdot x} (x^2)^{N\beta_1} = \rho(\beta_1,N) 
(k^2)^{-N\beta_1-d/2}  \ .   
\label{masslessint}  
\fqr 
The form in \rf{masslessint} is used with the derivatives in \rf{tensorprops} 
and \rf{propderivative} to deduce the tensor integrals; the expansion of 
the propagators is also to be included.  

The derivatives in \rf{propderivative} result in 

\bqr 
\int d^dx e^{ik\cdot x} \prod^a \partial_{\mu_j} \Delta^N 
\quad \sim \quad  
  \prod_i^{a_1} \eta_{\mu_{\rho(i)}\mu_{{\tilde\rho}(i)}} 
  \prod_j^{a_2} k_{\mu_\beta(j)}  
\label{ktensor}
\fqr 
The next set of derivatives act on the propagators.  The permutation 
set will be obtained as in the tree amplitude case, with direct sets of 
numbers.  

Expansions of the invariants are, 

\bqr 
(k_i+\ldots+k_j +{1\over N_1} (i\partial_k)^{-1})^{-2} 
\fqr 
\bqr = \sum_a (-1)^a
 \bigl( (k_i+\ldots+k_j)^2 + {1\over N_1} 
  ( k_i+\ldots+k_j)\cdot (i\partial_k)^{-1} \bigr)^{a} (-\partial_k^2)^{-a-1}  
\label{multiinvexp}
\fqr 
and in the two-particle case, 

\bqr 
(k_i+ (i\partial_k)^{-1})^2 =  
   (k_i\cdot {i\partial_k}^{-1} - ({\partial_k^2})^{-1}) 
\fqr  
\bqr 
= \sum_a (-1)^a
 \bigl( k_i\cdot (i\partial_k)^{-1} 
 \bigr)^{a} (-\partial_k^2)^{-a-1} \ . 
\label{twoparticleinvexp}
\fqr 
These expansions can be used to evaluate the tensor integrals that 
contain the propagators.  The third type of invariant is $(k_i+\ldots 
+k_j)^2= {1/N_1}(-\partial_k^2)$.  The sums in the terms \rf{multiinvexp} 
and \rf{twoparticleinvexp} then have to be performed.  The action on 
of the $\partial_k$ derivatives is taken on the scalar integral of the 
propagators, in \rf{ktensor}.  (The differential 
operators commute in x-space.)

The result for the product of propagators is 

\bqr 
\prod^b_\sigma t_{\sigma(i,)}^{-1,\sigma(,p)} = \prod_{j=1}
  \sum_{a_j} (-1) \bigl( q_j^2 + {1\over N_j} 
   q_j\cdot (i\partial_k)^{-1} \bigr)^{a_j} (\partial_k^2)^{-a_j-1} \ , 
\fqr 
\bqr 
= \prod_{j=1} \sum_{a_j=0}^\infty  
 \sum_{d_j=1}^{a_j} {a_j! \over d_j! (a_j-d_j)!} (-1) ~\bigl(q_j^2\bigr)^{d_j} 
  \bigl({i\over N_j} q_j \cdot \partial_k\bigr)^{a_j-d_j} 
 (\partial_k^2)^{-2a_j+d_j-1} 
\label{propderivatives}
\fqr 
with $t_i^{[p]}$ and $q_j$ the external portion of the external-internal 
momenta, with the former defined by $\sigma(i,p)$.  The product is 
over all invariants; when $q_j=0$ then only the 
$a_j=0$ term contributes.  There are a $b$ count of the derivatives, 
i.e. $b$ of the propagators with the internal momenta so that 
$b=\sum (a_j -d_j)$.

The combination of \rf{ktensor} and \rf{propderivatives} results in a 
series of derivatives as in \rf{ktensor}.  The derivatives from the 
tensor are  

\bqr 
\prod^a (i\partial_{k,\mu_i}) {1\over\partial_k^2} \ , 
\fqr 
and combine with the \rf{propderivatives}.  Then there are a total of $a+b$ 
derivatives in the numerator, which result in a tensor calculation; there 
are $b_2$ boxes in the denominator.

The result for the derivatives with $2{\tilde a}_1+{\tilde a}_2=a+b$ is, 

\bqr 
 \sum_{\rho,\tilde\rho,\tilde\beta} \sum_{2{\tilde a}_1+{\tilde a}_2}
  \prod_i^{a_1} \eta_{ \mu_{\rho(i)}\mu_{{\tilde\rho}(i)} } 
  \prod_j^{a_2} k_{\mu_\beta(j)} 
  \prod_l^{a_3} q_{ {\tilde\beta}(l),{\mu_{\tilde\beta}(l)} }
  (k^2)^{-N\beta-d/2-a_1/2-a_2+b_2} \ ,  
\fqr 
\bqr 
\times ~ \rho(\beta_1,N) 2^{a_1/2+a_2-b_2/2} 
 {\Gamma\bigl(-N\beta-d/2+b_2+1\bigr)\over 
  \Gamma\bigl(-N\beta-d/2-a_1/2-a_2+b_2+1\bigr)} 
  {\Gamma\bigl(N\beta+d/2-b_2\bigr)\over\Gamma\bigl(N\beta+d/2\bigr)} \ . 
\label{totaltensor}
\fqr 
The tensor is to be contracted with the function in \rf{propderivatives} from 
the propagator products and with vertex tensors in \rf{vertextensor} at 
nodes 1 and 2.  The propagators are $\eta_{\mu\nu}/p^2$.  The vertex tensors 
are discussed in the prior section and in \cite{Chalmers3}.  The tensor at 
node 2 used in the calculation contains the propagator expansions as at 
node 1, and the $\varepsilon$'s stripped and contracted with those at 
node 1.  The remaining momenta, in \rf{vertextensor}, if they are internal 
are integrated over.  

The loop integrations follow the form as illustrated in Figure 1. 
At each node the momenta $k_\sigma$, from $\sigma_b(m_i)$ to $\sigma_b(m_f)$, 
flow into the loop; these momenta define the nodal momentum $P_b$.  These 
tensors with the integral between the first two nodes result in, 

\bqr 
\rho(\beta_1,N) 2^{{\tilde a}_1/2+{\tilde a}_2-b_2/2} 
 {\Gamma\bigl(-N\beta-d/2+b_2+1\bigr)\over 
  \Gamma\bigl(-N\beta-d/2-{\tilde a}_1/2-{\tilde a}_2+b_2+1\bigr)} 
  {\Gamma\bigl(N\beta+d/2-b_2\bigr)\over\Gamma\bigl(N\beta+d/2\bigr)}
\fqr 
\bqr 
\times ~ \prod_{j=1} \sum_{a_j=0}^\infty  
 \sum_{d_j=1}^{a_j} {a_j! \over d_j! (a_j-d_j)!} (-1)  
  \bigl({i\over N_j}\bigr)^{a_j-d_j}
\label{propagatorexp}
\fqr 
in terms of the external line momenta $k_i$ and polarizations, 

\bqr 
(k^2)^{-N\beta-d/2-{\tilde a}_1/2-{\tilde a}_2+b_2}
 \prod_{i=1}^{c_1} k_{\kappa_1(i)}\cdot k_{\kappa_2(i)} 
 \prod_{j=1}^{c_2} \varepsilon(\beta_1(j))\cdot \varepsilon(\beta_2(j)) 
 \prod_{l=1}^{c_3} \varepsilon({\tilde\beta}_1(l)) 
   \cdot k_{{\tilde\beta}_2(l)} \ .
\label{tencontract}
\fqr 
The basis could be rewritten in terms of the propagator momentum $q_j$, 
but the tree amplitude forms and their divergences suggest to keep the 
same basis.  Due to the number of polarizations, $2c_2+c_3$ is equal 
to the number of external lines left of node 1; the number $c_1$ can 
be arbitrarily high due to the series in \rf{tencontract}.  

The correlated sets $\kappa_i$, $\beta_i$ ${\tilde\beta}_i$ (of 
dimension $c_1$, $c_2$, 
and $c_3$) are determined from: 1) the $n$-point scattering $\phi_n$ set 
of numbers, 2) expansion of the propagators that contain internal loop 
momentum, and 3) the inner product of the momenta and $\varepsilon_j$ 
with the integrated tensors \rf{totaltensor} including external momenta 
and metrics.  These sets are derived from numbers parameterizing the  
the classical tensors as in \rf{kappaterms}; these can be imported from 
the numbers labeling the propagators on the affiliated scalar diagram.  

These tensors of dimension $c_i$ in principle are determined 
from the well defined set of numbers such as the $\phi_n$ are, which label 
the $\phi^3$ diagrams, and from the node momenta $P_a=\sum k_i$ together 
with the polarization numbers entering into the two nodes.  

In a given diagram, illustrated in Figure 1, there are potentially external 
lines at all of the nodes $b_{\rm nodes}$.  The node numbers $b_{\rm nodes}$ 
range from $2$ to a maximum set by the coupling order; $n$-point tree 
amplitudes are of $g^{n-2}$ coupling order (without possible higher 
dimension terms in the classical action).  These couplings constrain the 
number of internal lines within the integrals and the nodes; $\sum_{b=1}^{b_{\rm nodes}} (n_b-2)$ is a fixed order, $n-2+2L$ for an $L$ loop $n$-point.  
The tree amplitudes with ghost lines also have to be included, in the 
covariant gauge with covariant $\varepsilon$'s (there are light-cone 
$\varepsilon$'s but the tree amplitudes are different).

The quantum numbers that specify the diagram in Figure 1 are the propagator 
labels $\sigma(i,p)$, or the equivalent $\phi_n$ numbers.  The node momentum 
indices $\sigma_b(m_i)$ to $\sigma_b(m_f)$ labels the external lines at each 
vertex.  The ordering of the lines is required also in the $\sigma$. 

The integration from nodes $1-2$, $2-3$, etc..., at a fixed order in their 
couplings, can be performed with a systematic implementation of the 
previous quantum numbers and node momentum.  The integration  
along the chain of vertices is influenced in a sequential manner because 
the tensor is altered sequentially.  The integration removes the 
propagators containing components of the node's momenta in the $j+1$ side 
of $j-j+1$, and the tensor (analagous to \rf{kappaterms}) then contains 
inner products which are non-tree like.  However, the individual tensors 
in \rf{kappaterms} do factor into the left node and right node with 
the use of metrics.  This property makes all of the integrals independent, 
at the computational cost of specifying the metrics.   

The result form of the $b_{\rm nodes}$ integrations contains the expansion 
of the external-internal momenta at each node , 

\bqr
 \rho(\beta_1,N) 2^{{\tilde a}_1/2+{\tilde a}_2-b_2/2} 
 {\Gamma\bigl(-N\beta-d/2+b_2+1\bigr)\over 
  \Gamma\bigl(-N\beta-d/2-{\tilde a}_1/2-{\tilde a}_2+b_2+1\bigr)} 
  {\Gamma\bigl(N\beta+d/2-b_2\bigr)\over\Gamma\bigl(N\beta+d/2\bigr)}
\fqr 
\bqr 
\times ~\prod_{b=1} \prod_{j=1} \Theta(N_j^{\sigma(i,p)}) \sum_{a_j=0}^\infty  
 \sum_{d_j=1}^{a_j} {a_j! \over d_j! (a_j-d_j)!} (-1)  
  \bigl({i\over N_j}\bigr)^{a_j-d_j}
\fqr 
and the kinematic factors, 
\bqr 
\prod_{b=1} (P_b^2)^{-n_b\beta-d/2-{\tilde a}_1/2-{\tilde a}_2+b_2} ~ 
\fqr 
\bqr 
\sum_{\kappa,\beta,{\tilde\beta}} 
 \prod_{i=1}^{d_1} k_{\kappa_1(i)}\cdot k_{\kappa_2(i)} 
 \prod_{j=1}^{d_2} \varepsilon(\beta_1(j))\cdot \varepsilon(\beta_2(j)) 
 \prod_{l=1}^{d_3} \varepsilon({\tilde\beta}_1(l)) 
   \cdot k_{{\tilde\beta}_2(l)} \ . 
\label{gaugeamplitudes}
\fqr  
The kinematic factors are due to various contractions at the nodes.  There 
are also the remaining propagators from the tree amplitudes, from the 
external momenta, 

\bqr 
\prod_{b=1}^{b_{\rm nodes}} \prod_{\sigma_b} {\tilde t}_{\sigma}^{-1} \ . 
\fqr 
The node momenta $P_a$ is used with the tree numbers (e.g. $\phi_n$) to 
restrict the invariants to those of the ${\tilde t}$ type and the $N_j$ 
numbers.  The $N_j$ count the number of internal momenta within an invariant 
$t_i^{[[p]}$, scaled by $1/(2-d)$; the latter is part of a tree used to 
define a node.  The number of internal lines is $\prod_{j=1}^{b_{\rm nodes}-1} 
N^{n_j}$, with $n_j$ counted to the right-hand side of the node.  All of this 
information is encoded in the numbers $\phi_n$, or an equivalent set such 
as $\sigma(i,p)$, that define the tree combinations.

The sets of $\beta$, $\tilde\beta$, and $\kappa$ depend on the tree 
contributions at the nodes and the external momenta configuration; there 
is likely a set theoretic definition of these groups of numbers.  

Last, the summation on the internal lines has to be performed in accord 
with the coupling conservation, i.e. $\sum (n_j-2)=n-2+2L$ and the sum 
over the node number and internal line number in \rf{gaugeamplitudes}.  
The external lines have to be permuted in line with a fixed color structure.

\vskip .2in 
\noindent{\it Gravity Amplitudes} 

The gravity amplitudes have the almost the same form as in 
\rf{gaugeamplitudes} except the kinematic factor has twice as many 
polarizations, 
\bqr 
\sum_{\kappa,\beta,{\tilde\beta},\beta',{\tilde\beta}'} 
 \prod_{i=1}^{d_1} k_{\kappa_1(i)}\cdot k_{\kappa_2(i)} 
 \prod_{j=1}^{d_2} \varepsilon(\beta_1(j))\cdot \varepsilon(\beta_2(j)) 
 \prod_{l=1}^{d_3} \varepsilon({\tilde\beta}_1(l)) 
   \cdot k_{{\tilde\beta}_2(l)} 
\fqr 
\bqr 
 \prod_{j=1}^{d'_2} {\bar\varepsilon}(\beta'_1(j))\cdot 
         {\bar\varepsilon}(\beta'_2(j)) 
 \prod_{l=1}^{d'_3} {\bar\varepsilon}({\tilde\beta}'_1(l)) 
   \cdot k_{{\tilde\beta}'_2(l)}
\fqr 
and the sums in $\kappa$, $(\beta$,$\tilde\beta)$,  $(\beta'$,$\tilde\beta')$ 
are altered.  The sums have a different dependence on $a_j$ and $d_j$.

In the derivation, the permutations on the external lines are completely 
symmetrized in the kinematics as there are no color quantum attached to 
the gravitons.  Also, mixed particle type scattering can be obtained, 
between gauge and gravity modes, and in general including varying spin types 
if the classical scattering is known. 

\vskip .2in 
\noindent {\it Concluding remarks}

The derivative expansion and its recursion is solved for both gauge 
and gravity theories.  Expressions for the amplitudes are obtained, 
and a closed form for them is given.  At any given order in derivatives, 
the operator's prefactor $f(\lambda)$ can be determined.   All the 
perturbative integrals are computed.

Three sets of similar tensor indices are required to make the 
expressions explicit, apart from a recursive integral.  Six sets are 
required for gravity.  The sets of indices are dependent on a number of 
parameters, and can be solved for recursively.  However, it is very 
suggestive that there is a group theory or topological determination 
in their determination, which is applicable to mixed particle scattering.

There is a power series in the momenta of the external lines.  The power 
series and its summation are likely to be relevant in certain kinematical 
regimes.  The tensor indices, and their possible topological determination, 
are relevant in this calculation.  

The amplitudes are written in terms of the polarizations, and thus they 
are not in a spinor helicity format.  There are further simplifications 
with the former.  Covariant tree amplitudes in the gauge and gravity 
theories are found in \cite{Chalmers1},\cite{Chalmers2}.
 
Knowledge of the coefficients in the perturbative sector can be used to 
find non-perturbative formalisms, i.e. strong coupling.  The global 
tensor determination, and its symmetry, is necessary for this.

\vfill\break


\begin{thebibliography}{99}

\bibitem{Chalmers4}
G. Chalmers, {\it Derivation of Quantum Field Dynamics}, physics/0503062. 

\bibitem{Chalmers5}
G. Chalmers, {\it Masses and Interactions of Nucleons in the
Derivative Expansion}, physics/0503110.

\bibitem{Chalmers6}
G. Chalmers, {\it Computing K3 and CY-n Metrics}, in preparation. 

\bibitem{Chalmer7}
G. Chalmers, {\it Comment on the Riemann Hypothesis}, physics/0503141. 

\bibitem{Chalmers8}
G. Chalmers, {\it Gauge Theories in the Derivative Expansion}, hep-th/0209086.

\bibitem{Chalmers9}
G. Chalmers, {\it Scalar Field Theory in the Derivative Expansion}, 
hep-th/0209075.

\bibitem{Chalmers10}
G. Chalmers, {\it M Theory and Automorphic Scattering}, Phys.\ Rev.\ D 
{\bf 64}:046014 (2001).

\bibitem{Chalmers11}
G. Chalmers, {\it On the Finiteness of $N=8$ Supergravity}, hep-th/0008162.

\bibitem{Chalmers12}
G. Chalmers and J. Erdmenger, {\it Dual Expansions of $N=4$ super Yang-Mills 
theory via IIB Superstring Theory}, Nucl.\ Phys.\ B {\bf 585}:517 (2000), 
hep-th/0005192.

\bibitem{Chalmers13}
G. Chalmers, {\it S and U-duality Constraints on IIB S-Matrices}, Nucl.\ 
Phys.\ B {\bf 580}:193 (2000), hep-th/0001190.

\bibitem{Chalmers3} 
G. Chalmers, {\it Quantum Scalar Field Theory Solution}, physics/0505018. 
 
\bibitem{Chalmers1} 
G. Chalmers, {\it Tree Amplitudes in Gauge and Gravity Theories}, 
physics/0504173.

\bibitem{Chalmers2} 
G. Chalmers, {\it Tree Amplitudes in Scalar Field Theories}, physics/0504219.

\end{thebibliography}
\end{document}